\documentclass[twocolumn,showpacs,preprintnumbers,amsmath,amssymb]{revtex4}
\usepackage{graphicx}
\usepackage{dcolumn}
\usepackage{bm}
\usepackage{color}

\begin{document}
\title{Entanglement can preserve the compact nature of the phase-space occupancy}

\author{A. M. C. Souza$^{1,2}$, P. Rap\v{c}an$^{2}$, and C. Tsallis$^{2,3,4}$}

\affiliation{
$^{1}$ Departamento de Fisica, Universidade Federal de Sergipe 49100-000, Sao Cristovao, Brazil.\\
$^{2}$ Centro Brasileiro de Pesquisas Fisicas and National Institute of Science and Technology for Complex Systems, Rua Dr. Xavier Sigaud 150, 22290-180,
Rio de Janeiro, Brazil\\
$^{3}$ Santa Fe Institute, 1399 Hyde Park Road, Santa Fe, NM 87501, USA\\
$^{4}$ Complexity Science Hub Vienna, Josefstadter Strasse 39, 1080 Vienna, Austria}

\date{\today}

\begin{abstract}
We study the one-dimensional transverse-field spin-1/2 Ising ferromagnet at its critical point.
We consider an $L$-sized subsystem of a $N$-sized ring, and trace over the states of $(N-L)$ spins,
with $N\to\infty$. The full $N$-system is in a pure state, but the $L$-system is in a statistical mixture.
As well known, for $L >>1$, the Boltzmann-Gibbs-von Neumann entropy violates thermodynamical extensivity,
namely $S_{BG}(L) \propto \log L$, whereas the nonadditive entropy $S_q$ is extensive for $q=q_c=\sqrt{37}-6 $,
namely $S_{q_c}(L) \propto L$. When this problem is expressed in terms of independent fermions,
we show that the usual thermostatistical sums emerging within Fermi-Dirac statistics can,  for $L>>1$, be indistinctively taken up to $L$ terms or up to $\log L$ terms.
This is interpreted as a compact occupancy of phase-space of the $L$-system,
hence standard BG quantities with an effective length $V \equiv \log L$ are appropriate and are explicitly calculated.
In other words, the calculations are to be done in a phase-space whose effective dimension is $2^{\log L}$ instead of $2^L$.
The whole scenario is strongly reminiscent of a usual phase transition of a spin-1/2 $d$-dimensional system,
where the phase-space dimension is $2^{L^d}$ in the disordered phase, and effectively $2^{L^d/2}$ in the ordered one.
\end{abstract}


\maketitle

The Boltzmann-Gibbs (BG) theory refers to ensembles, which constitute pillars of statistical mechanics \cite{book}.
The microcanonical ensemble, for example, is associated with the set of points in the phase-space in which one can choose a given total energy.
In this case, it is assumed a priori the ergodic hypothesis, in which the trajectories of the particles cover the hypersurface of the phase-space corresponding to that energy
in a time scale sufficient to carry out the measurements.
In many cases, it is not necessary for the system to cover the entire phase-space associated with the ensemble in question, but only a finite part of it, for instance half of it.
A typical example is usual phase transitions. Below a certain critical temperature, the system has a spontaneous symmetry breaking and effectively
occupies only half the phase-space. However, we may still assume the ergodic hypothesis in this half, and thus remain within the BG theory.

A more complex situation occurs in disordered glass-like systems \cite{palm}, in which the particles cover a small volume of the phase-space corresponding to a vanishing Lebesgue measure \cite{lebe}.
In this case, we can think in two situations: (i) the particles have trajectories that cover a compact subspace of the total phase-space, or, (ii) the particles have trajectories that do not cover the total phase-space,
at all relevant time scales, and it is not possible to identify a compact subspace.
In studies of conservative nonlinear dynamical systems, some examples of this latter situation has been found \cite{tirn}.
In this case, ergodicity might be broken in such a complex manner that the use of BG theory may be not legitimate.
Weak chaotic regimes have been found and the $q$-statistical generalization  \cite{tsal1} of the BG theory has emerged as an appropriate description.

Here, we analyze the nature of the phase-space occupancy as function of the entanglement. We focus on
the one-dimensional transverse-field spin-1/2 Ising ferromagnet at its zero-temperature critical point \cite{pfeu,sach}.
We consider an $L$-sized subsystem of a $N$-sized ring, and trace over the states of $(N-L)$ spins,
with $N\to\infty$. The full $N$-system is in a pure state, but the $L$-system is in a statistical mixture.
We show that the quantum entanglement becomes responsible for trapping part of the particles into not physically attainable energy states.
In this sense, particles do not cover the total volume of the phase-space. However, they do not destroy the nature of the phase-space ocuppancy, covering a {\it compact} subspace of the total phase-space. Therefore, the BG theory can continue to be legitimately used.
In this case, we can recover extensivity for physical quantities such as the entropy, which can linearly grow with the system size, by redefining the size of
the system.

Entanglement stands out as a key-feature in the mechanics underlying quantum phase transitions \cite{amic,its,lato,cala,isla,vida}.
A few years ago, Vidal {\it et al} \cite{vida} proposed an entanglement measure for pure states based on the
von Neumann entropy of the reduced density matrix accounting for a subset of the total system. This can be done by simply tracing out
the undesired external degrees of freedom. Entanglement undergoes a prominent increase in the vicinity of a critical point \cite{amic},
at which the von Neumann entropy acquires finite values while we approach the quantum phase transition point, where it diverges.
Fortunately, conformal field theory allows for an analytical description of the physical properties right at the transition point \cite{cala}.
It can be shown \cite{cala} that entanglement increases with the size of the subsystem $L$. Its divergence can be further associated to a given
universality class provided by conformal field theory \cite{holz}.

We stress that, when computing entanglement via the entropy, there is no need to impose conditions that usually follow from
their definition in order to establish a connection with thermodynamics.
However, here the entanglement entropy naturally assumes the role of thermodynamic entropy as well, by allowing for extensive thermodynamic variables corresponding to an effective volume of the system.

The one-dimensional transverse-field spin-1/2 Ising ferromagnet with $N$ sites is
described by the following Hamiltonian \cite{pfeu}
\begin{equation} \label{eq1}
\hat{H}= - \sum_{i=0}^{N-1} \left( \sigma_{i}^{x} \sigma_{i+1}^{x} + \lambda \sigma_{i}^{z} \right),
\end{equation}
where $\sigma_{i}^{\alpha}$ is the $\alpha$th Pauli matrix at site $i$ and $\lambda$ denotes
the magnetic field along the $z$ direction.
The Hamiltonian (\ref{eq1}) can be diagonalized by a Jordan-Wigner transformation, which
maps the spin chain onto a spinless fermionic system, followed up by a Bogoliubov linear transformation \cite{sach}.
The Hamiltonian then assumes the diagonal form
\begin{equation} \label{eq2}
\hat{H}= - \sum_{k} \left( \omega_{k} \hat{a}_{k}^{\dagger}
\hat{a}_{k} + \lambda \omega_{k} \right),
\end{equation}
where $\hat{a}_{k}$ are operators that obey the usual fermionic
anticommutation relations,  $k=-N/2,-N/2+1,...,N/2-1$ and
$\omega_{k}= \sqrt{ [\sin{(2\pi k/N)}]^{2} + [\cos{(2\pi k/N)} -
\lambda]^{2} }$.

The ground-state properties of this model strongly depend on $\lambda$.
A zero-temperature quantum phase transition occurs when $\lambda = 1$.
The ground-state behavior is further revealed by the interplay between entanglement and
the ground-state structure \cite{amic}.
Right at the critical point $\lambda = 1$ the spins are
mostly entangled and, in this case, it is possible to define a
proper entanglement witness which brings about signatures of
a quantum phase transition.

One of the most commonly-used entanglement measures for such a task
is the so-called entanglement entropy \cite{vida}. Given a pure state, it quantifies how much a given
subsystem, which can be properly described by a reduced density matrix,
is entangled with the remaining part. For a spin chain with $N$ sites, we obtain the state describing
a given block of $L$ spins $\rho_{L}$ by tracing out the subsystem of length $(N-L)$ of the overall density matrix
$\rho_{N}$. We have taken the thermodynamic limit ($N \rightarrow \infty$).

The von Neumann entanglement entropy reads \cite{vida}
\begin{equation} \label{s1}
S_{vN}(L,\lambda)= - Tr [ \rho_{L} \log ( \rho_{L})],
\end{equation}
the R\'enyi entropy \cite{fran,hast} reads
\begin{equation} \label{s2}
S_{\alpha}^R(L,\lambda) =  \frac{ 1 }{1-\alpha} \log [Tr
(\rho_{L})^{\alpha}]\,,
\end{equation}
and the $q$-entropy \cite{tsal1} reads
\begin{equation} \label{s3}
S_{q}(L,\lambda) =  \frac{ 1- Tr (\rho_{L})^{q} }{q-1}\,.
\end{equation}
Note that many other entanglement measures can be defined \cite{horo}.
Regardless of the choice though, all the relevant information is contained in the
reduced density matrix $\rho_{L}$.

Let us first discuss the entanglement properties when we are away from the critical point, that is $\lambda\neq1$ (recall that $L
\rightarrow \infty$). Using the mapping between the quantum $d=1$ model and the classical $d=2$ model, it is possible to express
$\rho_{L}$ as a product of density matrices account for an infinite number of uncorrelated spinless free fermions \cite{cala}.
The energy levels of the fermions are
\begin{equation} \label{epl}
\epsilon_{\lambda} (n) = \left\{
  \begin{array}{ll}
    (2n+1)\epsilon_{\lambda}, & \hbox{for} \; \lambda < 1 , \\
    2n\epsilon_{\lambda}, & \hbox{for} \; \lambda > 1 ,
  \end{array}
\right.
\end{equation}
with $n=0,1,2,...$ and
\begin{equation}
\epsilon_{\lambda} = \pi \frac{I(\sqrt{1-y^{2}})}{I(y)},
\end{equation}
where
\begin{equation}
I(y) = \int_{0}^{1} \frac{dx}{\sqrt{(1-x^{2})(1-y^{2}x^{2})}}
\end{equation}
is the complete elliptic integral of the first kind and $y=min[\lambda,\lambda^{-1}]$. Therefore, $\rho_{L \rightarrow
\infty}= \otimes_{n} \tilde{\rho}_{n}$, where
\begin{center}
\begin{equation} \label{rhol}
\tilde{\rho}_{n} = \frac{1}{ 1+e^{ -\epsilon_{\lambda} (n)} } \left(
\begin{array}{cc}
1 & 0  \\
0 & e^{-\epsilon_{\lambda}(n)}
\end{array}
\right).
\end{equation}
\end{center}

Once we have obtained the reduced density matrix, we can calculate
the von Neumann entropy for, say, $\lambda > 1$, using
\begin{equation} \label{s1a}
S_{vN}(\infty,\lambda) = \sum_{n=0}^{\infty} \left[ \log
\left(1+e^{-2n\epsilon_{\lambda}}\right) +
\frac{2n\epsilon_{\lambda}}{1+e^{-2n\epsilon_{\lambda}}} \right].
\end{equation}
In the vicinity of the critical point ($\lambda \rightarrow 1$),
we have that $\epsilon_{\lambda} \rightarrow 0$ and the sum above can be approximated by the integral
\begin{equation} \label{s1b}
S_{vN}(\infty,\lambda) \simeq \int_{0}^{\infty} dx \left[ \log
\left(1+e^{-2x\epsilon_{\lambda}}\right) +
\frac{2x\epsilon_{\lambda}}{1+e^{-2x\epsilon_{\lambda}}} \right]
\end{equation}
\begin{equation} \label{s1c}
S_{vN}(\infty,\lambda) \simeq \frac{\pi^{2}}{12\epsilon_{\lambda}}
\rightarrow \infty.
\end{equation}
We can unveil the behavior of the Renyi and of the $q$-entropy by a similar procedure \cite{jin,caru}.

\begin{figure}[h!]
\centering
\includegraphics[width=8cm,angle=0]{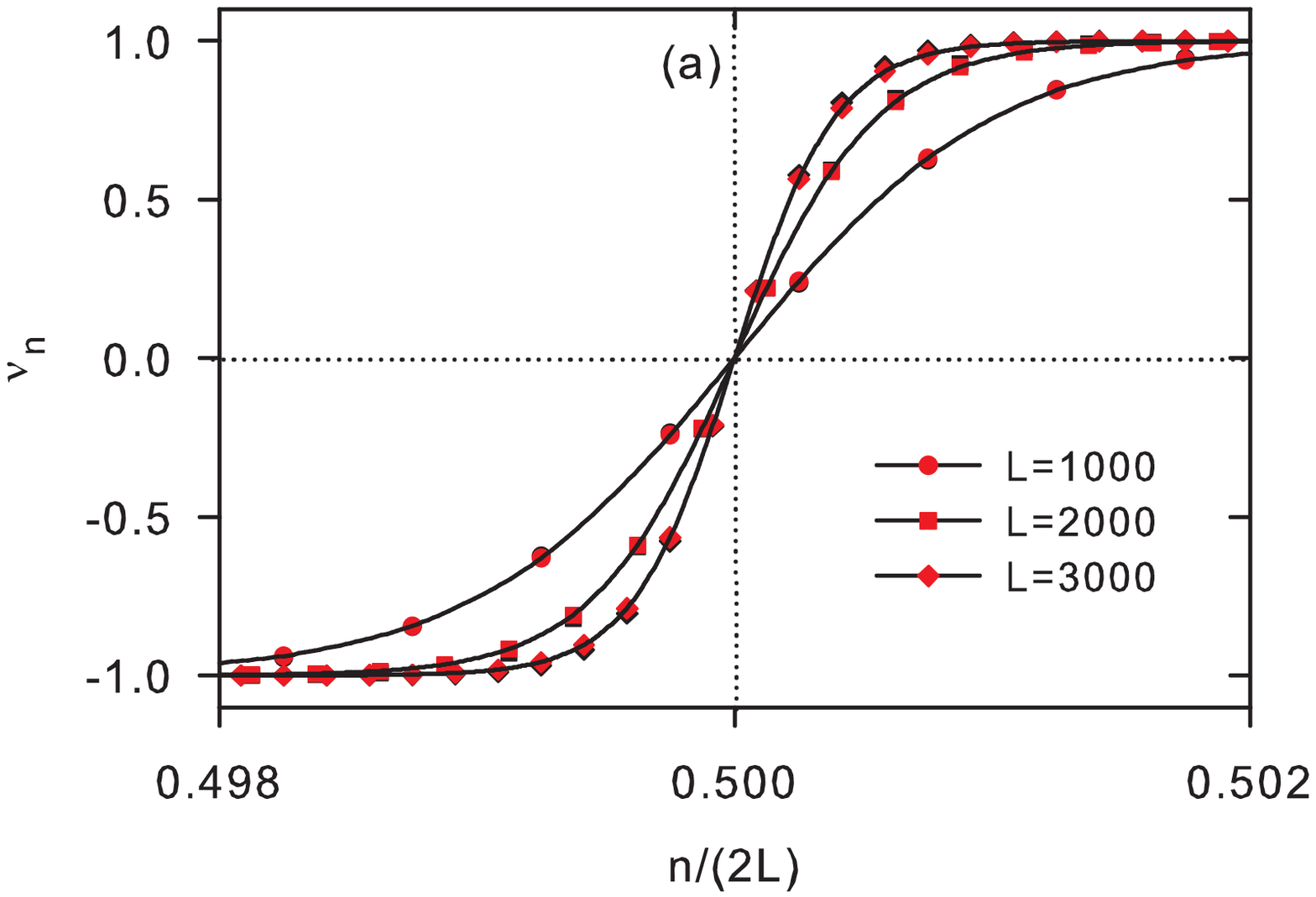}
\includegraphics[width=8cm,angle=0]{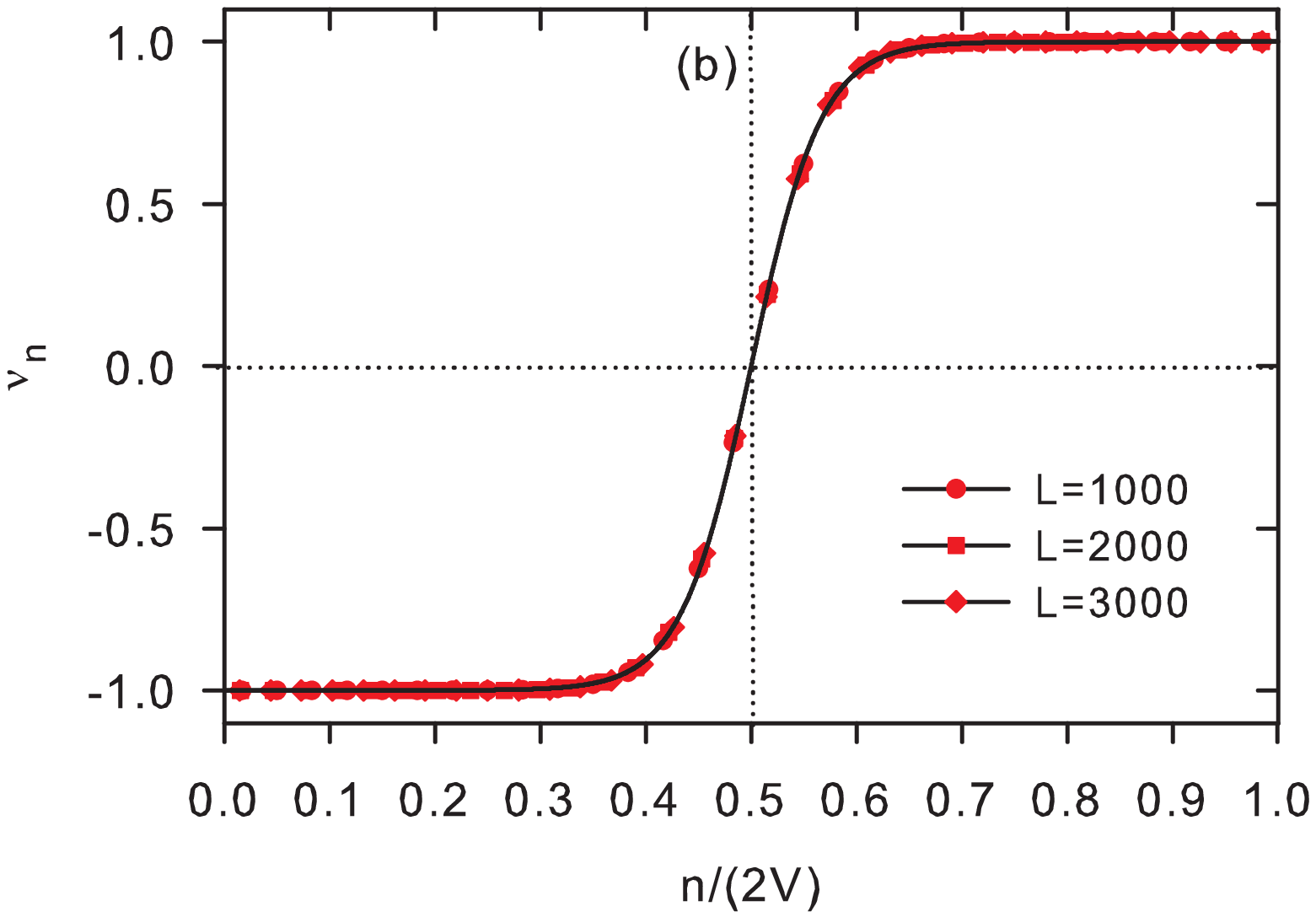}
\caption{(a) Imaginary part of the eigenvalues of $\Gamma_{L}$ as a
function of $n/2L$ obtained by numerical diagonalization for
$L=1000$, $2000$ and $3000$. Black symbols represent the numerical
results and the red ones denote the theoretical result $\nu_{n}=\tanh
[(2n+1)\epsilon_L /2]$, where $\epsilon_L$ is given by Eq.
(\ref{epl3}). (b) The same as in panel (a) but the abscissa is $n/2V$ with $V=1.227
\ln_{2}(L)$. Curves are guides to the eye. } \label{eigen}
\end{figure}

At the critical point, a similar analysis can be carried out. By considering now an $L$-sized  subsystem,
its reduced density matrix $\rho_{L}$ is obtained from the following matrix \cite{vida,lato}
\begin{equation}
\Gamma_{L} = \left(
  \begin{array}{cccc}
    \Pi_{0} & \Pi_{1} & \cdots & \Pi_{L-1} \\
    \Pi_{-1} & \Pi_{0} &  & \vdots \\
    \vdots &  & \ddots & \vdots \\
    \Pi_{1-L} & \cdots & \cdots & \Pi_{0} \\
  \end{array}
\right),
\end{equation}
where
\begin{equation}
\Pi_{l} = \left(
\begin{array}{cc}
0 & \frac{-4}{\pi (2l+1)} \\
\frac{-4}{\pi (2l-1)} & 0
\end{array}
\right).
\end{equation}
An orthogonal matrix transforms $\Gamma_{L}$ into a block-diagonal matrix corresponding to purely
imaginary eigenvalues $\pm i\nu_{n}$ ($n=0,..,L-1$). In Fig. \ref{eigen}(a), we show the imaginary part
of the eigenvalues of $\Gamma_{L}$ obtained through straightforward numerical diagonalization for various block sizes.
The analytical outcome for $\nu_{n}$ reads $\nu_{n}=\tanh [(2n+1)\epsilon_L /2]$,
where $\epsilon_L$ will be obtained later on. The $2^{L}$ eigenvalues of $\rho_{L}$ are given by
\begin{equation}
 \mu_{x_{1}x_{2}...x_{L}} = \prod_{n}  \frac{1+(-1)^{x_{n}}\nu_{n}}{2}
\end{equation}
where $x_{n}=0,1$ $\forall n$.

Analogously to our previous discussion for the $\lambda\neq1$ case, at the critical point the
model can also be mapped onto a system featuring spinless free fermions and thus
$\rho_{L}= \otimes_{n} \hat{\rho}_{n}$, where $\hat{\rho}_{n}$ has a similar form to that of Eq. (\ref{rhol}),
with $\epsilon_{\lambda}$ being replaced by $\epsilon_{L}$. For large $L$, we can write the energy spectrum as
\begin{equation} \label{epl2}
\epsilon_{L} (n) = (2n+1)\epsilon_{L} \;\;\;\; n=0,1,...,L-1.
\end{equation}
Similar spectrum was obtained by Peschel \cite{pesc} working with a matrix which commutes with $\Gamma_{L}$.
We obtained $\epsilon_{L}$ theoretically, observing that $\epsilon_{L} \rightarrow 0$ for $L \rightarrow \infty$,
so that, analogously to Eqs. (\ref{s1a}), (\ref{s1b}), and (\ref{s1c}), we can write
\begin{equation}
S_{vN}(L,1) \simeq \frac{\pi^{2}}{12\epsilon_{L}}.
\end{equation}
Using that \cite{cala,vida}
\begin{equation}
S_{vN}(L,1) \simeq \frac{1}{6} \log (L),
\end{equation}
we obtain
\begin{equation} \label{epl3}
\epsilon_{L} = \frac{\pi^{2}}{2\log (L)}.
\end{equation}
Our analytical results for the eigenvalues of $\Gamma_{L}$ are
represented in Fig. \ref{eigen}(a), which are in excellent agreement
with the numerical outcomes. Our numerical proof was made for finite
$L$, and in these cases, for high energies, we obtained a nonlinear
spectrum in $n$. This nonlinearity decreases with increasing $L$ and
strongly suggests a linear spectrum in $n$ for infinite $L$. Let us
stress that the region of high energies are physically irrelevant as
we shall see below.

We can think of the problem not solely for the purpose of
entanglement analysis, but also regarding the spin block as a
physical system of interest by itself. Thus, it becomes relevant to
discuss its thermodynamic properties, which is carried out in what
follows. We have a system of $L$ free fermions whose Hamiltonian
reads as
\begin{equation} \label{hamilf}
\hat{H}_{L} = E_L \sum_{n=0}^{L-1} (2n+1) \hat{c}^{\dag}_{n}
\hat{c}_{n},
\end{equation}
where $\hat{c}^{\dag}_{n}$ ($\hat{c}_{n}$) are the creation (annihilation) fermionic operators at site $n$ for a
one-dimensional lattice. $E_{L} = \epsilon_L \epsilon_0$, where both
$\epsilon_{0}$ and $\hat{H}_{L}$ have the dimension of energy.

The Hamiltonian (\ref{hamilf}) represents tightly-bounded electrons in a uniform electric field \cite{fuku,sait}. This model
has been extensively studied, both on theoretical and experimental grounds (see, e.g., \cite{mend,guti}).
Our case, however, embodies the limit of localized atomic electrons, where nearest-neighbor hopping is
neglected. In this extreme limit, the equidistant energy levels are identified as Stark ladders \cite{kane}.
The concept of Stark ladder was put forward by Wannier \cite{wann}
and confirmed experimentally in several setups, for instance, in GaAs-GaAlAs
superlattices subjected to electric fields \cite{mend} and in an elastic-rod
apparatus \cite{guti}.

The thermodynamic properties of the free fermions at temperature $T$
are determined from the partition function of the canonical ensemble
\begin{equation}  \label{fpa}
Z(L) = Tr  [ e^{-\beta \hat{H}_{L} }] = \prod_{n=0}^{L-1}
(1+e^{-\beta (2n+1)E_{L}}),
\end{equation}
where $\beta=1/(k_{B}T)$. We obtain the Helmholtz free energy \begin{equation} \label{he1}
F (L) = - \frac{1}{\beta} \ln [Z(L)]  = - \frac{1}{\beta}
\sum_{n=0}^{L-1} \ln [1+e^{-\beta (2n+1)E_L}]
\end{equation}
and the internal energy
\begin{equation} \label{in1}
U (L) = - \frac{\partial}{\partial\beta} \ln [Z(L)]  =
\sum_{n=0}^{L-1} \frac{ (2n+1)E_L }{1+e^{\beta (2n+1)E_L}}.
\end{equation}
As in Eq. (\ref{s1b}), the above sums can be approximated by integrals and we obtain
\begin{equation} \label{he2}
F (L) \simeq - \frac{\pi^{2}}{24 E_{L}\beta^{2}} = - \frac{1}{
12\beta^{2} \epsilon_{0}} \log  (L)
\end{equation}
and
\begin{equation} \label{in2}
U (L) \simeq \frac{\pi^{2}}{24 E_{L}\beta^{2}} =
\frac{1}{12\beta^{2} \epsilon_{0}} \log  (L) \,.
\end{equation}
Consequently, it becomes straightforward to obtain the entropy, which reads
\begin{equation} \label{s1d}
S (L) = \frac{1}{T} [U(L) - F(L)] \simeq \frac{k_{B}}{6\beta
\epsilon_{0}} \log  (L).
\end{equation}
One can recover the entanglement entropy by assuming $\beta
\epsilon_{0}=1$ and thus $S_{vN}(L,1) \simeq \frac{k_{B}}{6}
\log  (L)$. We can also write the R\'enyi \cite{fran} and
$q$-statistics entropies \cite{caru} as
\begin{equation} \label{s2d}
S_{\alpha}^R(L,1) \simeq  \frac{(\alpha+1)}{12\alpha}
\log  (L)
\end{equation}
and
\begin{equation} \label{s3d}
S_{q}(L,1) \simeq  \frac{ L^{(\frac{1}{q}-q)\frac{1}{12}} - 1
}{1-q},
\end{equation}
respectively.

Note that the $q$-entropy can be used by satisfying the requirement
of extensivity, i.e., $(\frac{1}{q}-q)\frac{1}{12}=1$, hence  $q_{c}=\sqrt{37}-6 \simeq 0.08$ \cite{caru}. In this case,
\begin{equation} \label{s4d}
S_{q_c}(L,1) \simeq L,
\end{equation}
and the desired thermodynamic extensivity is recovered.

We now make a crucial observation: this system has an effective
number of unattainable physical energy states  \cite{lato}, characterized by density matrices following
\begin{equation} \label{rhol2}
\hat{\rho}_{n} \simeq \left(
\begin{array}{cc}
1 & 0  \\
0 & 0
\end{array}
\right).
\end{equation}
Another way to put this is by thinking that we have a set of free fermions
frozen in the ground state that depends on the manner through which entanglement was established in the original problem.
Therefore, since only a part of the free fermions becomes thermodynamically
accessible, this is the very subset on which we build up our analysis.
For the number of accessible free fermion states versus the block size $L$,
for $L > 300$, we obtain numerically
\begin{equation} \label{lmax}
\tilde{L} = 1.227 \log L + 2.88  \;\;\; (r^{2}=0.999999).
\end{equation}
Those coefficients depend on the chosen numerical precision, which we set to $10^{-6}$ for real numbers.
However, it definitely has no influence on the qualitative features we point out next.

Fig. \ref{eigen}(b) shows $\nu_{n}$ [also featured in Fig. \ref{eigen}(a)]
now as a function of $n/(2V)$, where  $V\equiv 1.227 \log L$.
We can observe a data collapse at which $\nu_{n}$ is
independent of $L$.

Further numerical analysis yields the conclusion that the
entire physical behavior of the subsystem composed by $L$ free
fermions can be completely evaluated by considering only the first $V$
particles. This allows us to write
\begin{equation} \label{hamilv}
\hat{H}_{V} = E_L \sum_{n=0}^{V} (2n+1) \hat{c}^{\dag}_{n}
\hat{c}_{n}.
\end{equation}
Using this expression, we confirm that the results of
Eqs. (\ref{he2}), (\ref{in2}), (\ref{s1d}), (\ref{s2d}), and (\ref{s3d}) are precisely the same.

The thermodynamic properties are extracted from the free energy
\begin{equation} \label{he3}
F (T,V) = - \frac{k_{B}^{2}T^{2}}{12 \epsilon_{0}} V,
\end{equation}
such that
\begin{equation} \label{ent3}
S_{BG} (T,V) = - \left( \frac{\partial F}{\partial T} \right)_{V} = \frac{k_{B}^{2}T}{6 \epsilon_{0}} V
\end{equation}
and
\begin{equation} \label{ene3}
U (T,V) = F+TS = \frac{k_{B}^{2}T^{2}}{12 \epsilon_{0}} V
\end{equation}
are extensive thermodynamic quantities. For completeness, we can also define the intensive quantity
\begin{equation} \label{pre3}
P (T,V) = - \left( \frac{\partial F}{\partial V} \right)_{T} = \frac{k_{B}^{2}T^{2}}{12 \epsilon_{0}},
\end{equation}
so that we can write $U=PV$. All the above expressions are consistent with standard thermodynamics.

The entanglement behavior of the system mandates that only a given part of energy states is thermodynamically relevant.
As a consequence, the standard BG quantities are associated with an effective length $V \equiv \log L$, and the phase-space has an effective dimension $2^{\log L}$ instead of $2^L$.
The effective number of microstates grows with $L$ as a power-law, in variance
with the exponential growth corresponding to standard nonentangled systems.

The above analysis suggests a scenario where the physical systems are essentially grouped into three classes, in terms of their phase-space occupancy, ergodicity and Lebesgue measure, namely
(i) ergodicity occurs in the entire  phase-space or in a {\it compact} subspace whose Lebesgue measure remains different from zero in the thermodynamic limit;
(ii) ergodicity occurs only in a {\it compact} subspace whose Lebesgue measure vanishes in the thermodynamic limit; and
(iii) ergodicity does not occur, the trajectories covering a {\it noncompact} subspace whose Lebesgue measure vanishes in the thermodynamic limit (typically an hierarchical structure like a multifractal).
For each class, there is an appropriate statistical mechanics. Typical examples of the first class are physical systems with or without usual phase transitions.
The BG theory perfectly describes this class and the von Neumann/Boltzmann entropy is an extensive thermodynamic quantity.
For systems that fall in the second class, we exhibit in the present work  how the BG theory can still be used.
Here, we can find a particular value of $q$ such that the $q$-entropy satisfies the requirement of extensivity within the {\it total} volume, while the von Neumann/Boltzmann entropy is an extensive thermodynamic quantity
within an appropriate {\it effective} volume. Some of the systems exhibiting the area-law \cite {herd,cala2} for the entropy might also belong to this class.
For the third class, we do not expect the use of the BG theory to be legitimate. This is the case for say systems with long-range interactions,
for which theories such as $q$-statistics have been satisfactorily applied \cite{tsa_fp,tsal2,tsal3}.

{\it Acknowledgements} We thank G. M. A. Almeida, L. J. L. Cirto and
E. M. F. Curado for useful remarks, and Y. N. Fern\'{a}ndez, K.
Hallberg and A. Gendiar for very fruitful discussions at early
stages of this effort. We also acknowledge partial financial support
from the John Templeton Foundation-USA (Grant n\textordmasculine
53060) and from the Brazilian agencies CNPq and CAPES.

\end{document}